\documentclass[a4paper,fleqn,usenatbib]{mnras}

\usepackage{txfonts}

\usepackage[T1]{fontenc}
\usepackage{ae,aecompl}

\usepackage{graphicx}	

\title[Far Side-lobe Source Noise for SKA-LOW]{Analysing the impact of far-out side-lobes on the imaging performance of the SKA-LOW telescope}

\author[B. Mort et al.]{
Benjamin Mort,$^{1}$\thanks{E-mail: benjamin.mort@oerc.ox.ac.uk}
Fred Dulwich,$^{1}$
Nima Razavi-Ghods,$^{2}$
Eloy de Lera Acedo,$^{2}$
\newauthor
and Keith Grainge$^{3}$
\\
$^{1}$Oxford e-Research Centre, University of Oxford, 7 Keble Road, Oxford OX1 3QG, UK\\
$^{2}$Cavendish Astrophysics, University of Cambridge, JJ Thomson Avenue, Cambridge CB3 0HE, UK\\
$^{3}$School of Physics and Astronomy, University of Manchester, Alan Turing Building, Manchester M13 9PL, UK
}

\date{Accepted XXX. Received YYY; in original form ZZZ}

\pubyear{2016}

\begin{document}
\label{firstpage}
\pagerange{\pageref{firstpage}--\pageref{lastpage}}
\maketitle

\begin{abstract}
The Square Kilometre Array's Low Frequency instrument (SKA-LOW) will be the most sensitive aperture array ever used for radio astronomy, and will operate in the under-sampled regime for most of the frequency band where grating-lobes pose particular challenges. To achieve the expected level of sensitivity for SKA-LOW, it is particularly important to understand how interfering sources in both near and far side-lobes of the station beam affect the imaging performance. We discuss options for station designs, and adopting a random element layout, we assess its effectiveness by investigating how sources far from the main lobe of the station beam degrade images of the target field. These sources have the effect of introducing a noise-like corruption to images, which we call the Far Side-lobe Source Noise (FSSN). Using OSKAR, a GPU-accelerated software simulator, we carried out end-to-end simulations using an all-sky model and telescope configuration representative of the SKA-LOW instrument. The FSSN is a function of both the station beam and the interferometric point spread function, and decreases with increasing observation time until the coverage of the aperture plane no longer improves. Using apodisation to reduce the level of near-in side-lobes of the station beam had a noticeable improvement on the level of FSSN at low frequencies. Our results indicate that the effects of picking up sources in the side-lobes are worse at low frequencies, where the array is less sparse.
\end{abstract}




\section{Introduction}
\label{sec:intro}
The Square Kilometre Array (SKA)\footnote{http://www.skatelescope.org}, with its large collecting area and low receiver noise, has the potential to provide high fidelity, high dynamic range images over wide fields of view. In order to achieve its full potential, however, every aspect of the system will need to be analysed to unprecedented levels of detail. For the SKA's Low Frequency instrument (SKA-LOW), which will operate nominally in the 50-350~MHz frequency band and consist entirely of phased array antennas, these requirements present significant challenges. Signals from the hundreds of fixed elements that make up each station will be combined electronically to form beams in the required direction. This leads to station beams that exhibit complicated behaviour as a function of frequency and scan angle, and hence the beams will change significantly over the course of an observation. As a phased array telescope operating over such a wide bandwidth, the configuration of the antennas in each station plays an important role in defining station side-lobe profiles, which have a direct impact on the performance of the instrument as an interferometer. For SKA-LOW, the spatial Nyquist frequency is typically around 80~MHz \citep{Turner2015}, meaning that for the majority of the SKA-LOW band, the phased arrays are under-sampling the incoming wave front. Sparse arrays give rise to grating lobes, which exhibit different behaviour for regular and irregular type arrays, but in either case the presence of grating lobes implies that the beams formed at higher frequencies will be more sensitive to radiation from sources away from the direction of interest \citep{Razavi2012}. The spilled power detected from sources in the side-lobes or grating lobes of the station beam has the effect of contaminating observations made of the target field, and thus will limit the image dynamic range if it cannot be removed adequately.

While SKA pathfinder telescopes such as the Low Frequency Array (LOFAR) \citep{Haarlem2013} and the Murchison Widefield Array (MWA) \citep{Tingay2013} can test the use of aperture arrays at low frequencies, data sets from real instruments typically contain additional unwanted corruptions that can be difficult to deal with, and these corruptions can mask other important effects. Simulations provide clean environments that allow us to understand and improve the SKA-LOW design. Notably, because the sensitivity of the SKA is expected to be so much higher than that of the pathfinders, effects that are negligible on those instruments may become critical for SKA.

In the first instance, a great deal may be learnt by attempting to optimise the configuration of the antenna positions and their weighting scheme within each station and by analysing the performance of individual stations \citep{Razavi2012,Clavier2014}. However, this analysis must be extended to end-to-end simulations of the interferometer as a whole in order to understand how these station beam side-lobes can affect the resulting images \citep{Smirnov2011a,Smirnov2011b}.

End-to-end simulations of interferometers as large as SKA-LOW are extremely computationally intensive. Recent developments in simulation software, such as OSKAR\footnote{http://oskar.oerc.ox.ac.uk}, which evaluates the radio interferometer measurement equation \citep{Hamaker2006} using graphics processing units (GPUs) to give at least an order of magnitude improvement in performance over traditional simulation tools, make it possible to run large-scale, full-sky simulations of aperture arrays on reasonable timescales.

In this paper, we present the results of an investigation in which we generated simulated visibilities using a large number of interfering sources far from the main lobe of the primary (station) beam of an SKA-like telescope consisting entirely of aperture arrays. These sources have the effect of introducing a noise-like corruption to images, known as the Far Side-lobe Source Noise (FSSN). Whilst many components of system noise behave like Gaussian random variables, and therefore tend to decrease as observation length increases, FSSN is due to a multitude of side-lobes from real sources that will not tend to zero simply by integrating for an infinite time. The amount of power introduced into the field is proportional to the side-lobe level of the point spread function (PSF) and the apparent flux $I_a$ of each interfering source $s$ in the element beam:

\begin{equation}
{\rm FSSN} = \sum_{s} I_a \cdot {\rm PSF}
\end{equation}

\noindent where all of these quantities are a function of source position, time and frequency. The level of FSSN is therefore a function of both the aperture plane (UV) coverage and the overall ability of the interferometric cross-power station beam to act as an effective spatial filter on the sky.

The level of FSSN can be reduced by removing the brightest and closest interfering sources using, for example, the CLEAN algorithm \citep{Hogbom1974} or other source-peeling techniques if the emission from those sources can be modelled accurately. Removing the FSSN contribution of a source in this way requires excellent knowledge of the instrument to obtain both the apparent flux of the source, and an accurate PSF at the source position. Sparse aperture array stations, which naturally have more complex structure in their beam profiles than those from dish antennas or other filled apertures, give rise to apparent source fluxes that are higher and more variable away from the field of view. In practice, even if the instrument can be characterised very accurately, which is increasingly difficult far from the centre of the primary beam, the high computational cost will eventually limit the number of sources that can be removed using this procedure, and a residual FSSN component will remain. For a very sensitive instrument like the SKA, which will have exceptionally low thermal noise characteristics, FSSN may present a limiting factor in the noise performance for any science experiment that aims to remove all sources of contamination due to weak foregrounds.

We have chosen to use FSSN as it proves a practical metric that tracks the performance of the aperture array station beams. Since the station beam has a considerable effect on the apparent source flux, apodisation techniques can be used to improve the quality of the spatial filter the station beam provides. A sufficient level of control over the station beam side-lobes will be necessary for the SKA-LOW telescope, and this is described in Section~\ref{sec:designs}. In Section~\ref{sec:sims}, we describe the simulations we performed to study the level of FSSN as a function of time and frequency using a representative SKA-LOW telescope model. In our analysis, we have also included a frequency of 650 MHz as a possible extended frequency band for SKA-LOW. We present the results of these simulations in Section~\ref{sec:results}. Finally, we discuss the implications of our results and present our conclusions in Section~\ref{sec:conclusions}.

%
\section{SKA-LOW Array Configuration Designs}
\label{sec:designs}
The configuration of the antennas in an aperture array has been an active topic of research in radio astronomy for the past few years \citep{Braun2006,Razavi2012}. Phased array antenna systems may be implemented with regular as well as irregular geometries \citep{Cappellen2006}.  The former is well understood in the literature \citep[e.g.][]{Hansen1998}, especially in regards to the dense (separation $\leq \lambda/2$) and sparse (separation $> \lambda/2$) regimes where grating lobes are present.

Regular type arrays consist of not only typical regular or hexagonal lattices, but also aperiodic array types, which can range from the Danzer or Penrose tiles to the more exotic snowflake configurations.  What is apparent for all such arrays is the formation of grating lobes when operating in the under-sampled regime.  In this regime, a regular array would result in grating lobes, whilst below this frequency the array behaves very much like a continuous aperture. At a higher operating frequency where the array is sparser, the number of grating lobes increases and they also move closer to the main beam.  Modern wideband array antennas, such as those proposed for the SKA, may be implemented with irregular/random configurations to decrease grating lobes when working in the sparse regime \citep{Razavi2012}. However, the power, which would form a grating lobe in a regular array, is redistributed into a broad region of weaker irregular side-lobes, which are visible at all wavelengths.  This may reduce the quality of the beam and can have an impact on the calibration of the instrument.

For SKA-LOW, in order to meet the sensitivity requirements \citep{Turner2015}, a sparse configuration is needed to meet constraints of cost.  Due to the galactic synchrotron emission dominating at the low frequencies, in order to maximise the array sensitivity without increasing the number of elements, enough space needs to be allocated to each antenna in the array so that mutual coupling does not limit their effective aperture \citep{deLera2011a,deLera2015}.  Furthermore, in order to deliver maximum brightness sensitivity, the filling factor of the SKA-LOW core must be as high as possible \citep{Mellema2013}.  In the SKA-LOW core, all the individual antenna elements contribute to capture the relevant Fourier modes in a given angular scale, therefore a higher filling factor will maximize the information collected from the sky.  This imposes a filling factor requirement, which sets some limits on the distance between elements and the required footprint per antenna.  In the current design \citep{Turner2015}, a trade off average spacing between elements of $\sim1.9$ m has been chosen.

From the electromagnetic perspective, when discussing disconnected arrays as are proposed for SKA-LOW, antenna regularity will result in in-band resonances due to the mutual coupling of antennas, which cannot be avoided in wide band systems with bandwidths larger than 4:1. It can be observed that for the current SKA Log-periodic Antenna (SKALA), multiple in-band resonances appear when the antennas are placed in a regular configuration with $\lambda/2$ spacing at 100~MHz \citep{deLera2015}.

\begin{figure*}
\centering
\includegraphics[width=7.0in]{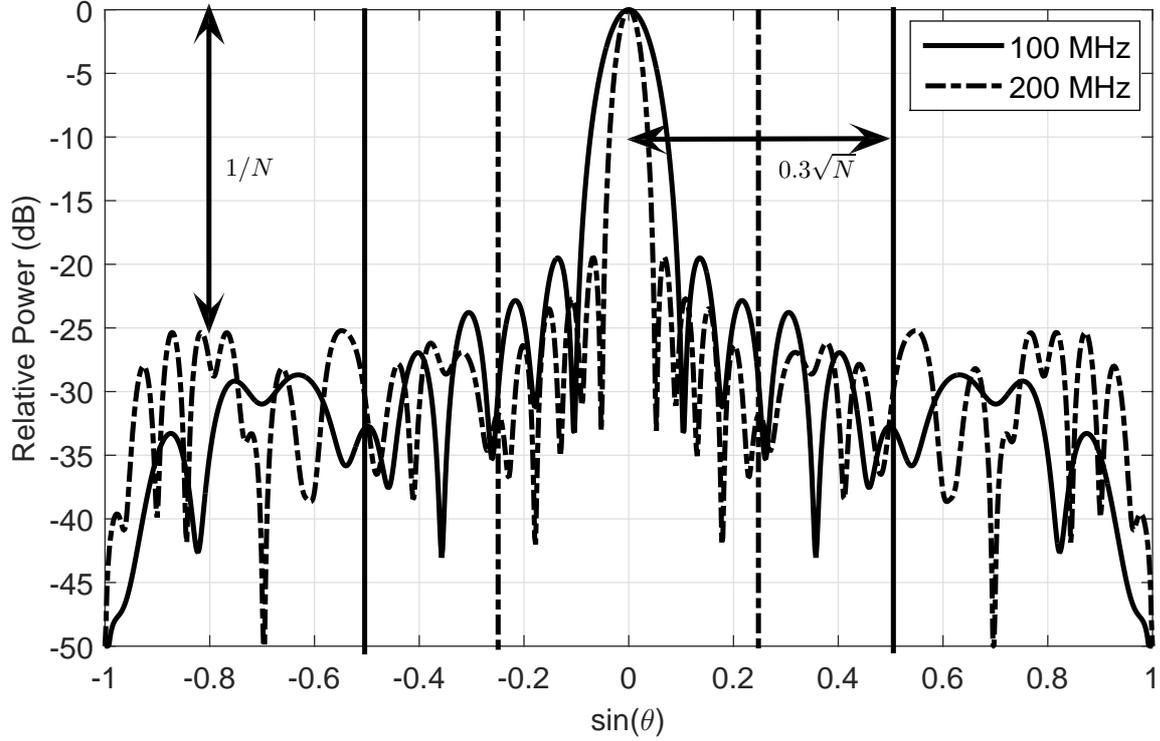}
\caption{A cut of the typical expected SKA-LOW station beam incorporating the SKALA response at 100~MHz and 200~MHz. The dotted lines show the transition region from dense to sparse.}
\label{fig1}
\end{figure*}

For these reasons, it has been assumed that an irregular type array must be implemented, i.e. one where the antenna positions are randomised.  In a truly random array, the grating lobes would exist in the form of irregularly distributed side-lobes, which can be analysed by separating the array beam into so-called ``coherent'' and ``non-coherent'' regions.  This terminology is described in detail in \citet{Clavier2014}, where we estimate the boundary of this coherent to non-coherent region to appear after $\sim 0.3\sqrt{N}$ side-lobes from the main beam, and the average level of side-lobes to be given by $\sim 1/N$, where $N$ is the number of antennas.  For the SKA-LOW telescope, where each station contains nominally 256 antennas \citep{Turner2015}, this transition would be $\sim 5$ side-lobes from the main beam (Figure~\ref{fig1}) with the typical widths of side-lobes being $\lambda/2D$, where $D$ is the diameter of the array. A higher operating frequency and thus more sparseness in a random array will result in this transition region appearing nearer the main beam in absolute terms. It is important to note that the side-lobe profile of each station may be different, since each station could use a different pseudo-random configuration.  We explore this added advantage later.

When examining the effects of mutual coupling for a randomised array, studies have shown \citep{Gonzalez2011,deLera2011b} that there is an averaging effect, which improves the more ``randomised'' the positions of the antennas become and the greater the number of antennas in a given station.  In \citet{Gonzalez2011} we show through EM simulations that the average of the embedded element patterns approaches the single (isolated) element pattern, which is even apparent for only 256 antennas in a station \citep{deLera2011b}.  Therefore, a real advantage of randomisation of antenna elements is in the modelling of the SKA-LOW telescope, since it is now possible to accurately examine the station response to first order using only a single element pattern simulated in standard electromagnetic software simulation packages, such as Computer Simulation Technology (CST) \footnote{http://www.cst.com}. This is the approach taken for the simulations presented in this paper.  It is important to note that during calibration, the inclusion of an accurate account for these mutual coupling effects is critical in order to achieve dynamic ranges better than 60~dB \citep{deLera2013}.
It has been demonstrated by LOFAR HBA, which has regular phased-array stations, that by rotating the antenna positions, the grating lobes appear at different places on the sky.  This results in the strong grating lobes of one station appearing where the response of the second station is low with the result that on multiplication, they average down dramatically in the power beam \citep{Wijnholds2011}.  For the SKA-LOW, we would also benefit from each array having a different random configuration and thus again resulting in the side-lobes of the station beam appearing at different positions on the sky, meaning in the interferometric sense the contribution of faraway sources in the side-lobes can appear reduced.

\subsection{Array layout and optimisation for control of side-lobes}
\label{sec:arrays}
\begin{figure*}
\centering
\includegraphics[width=3.3in]{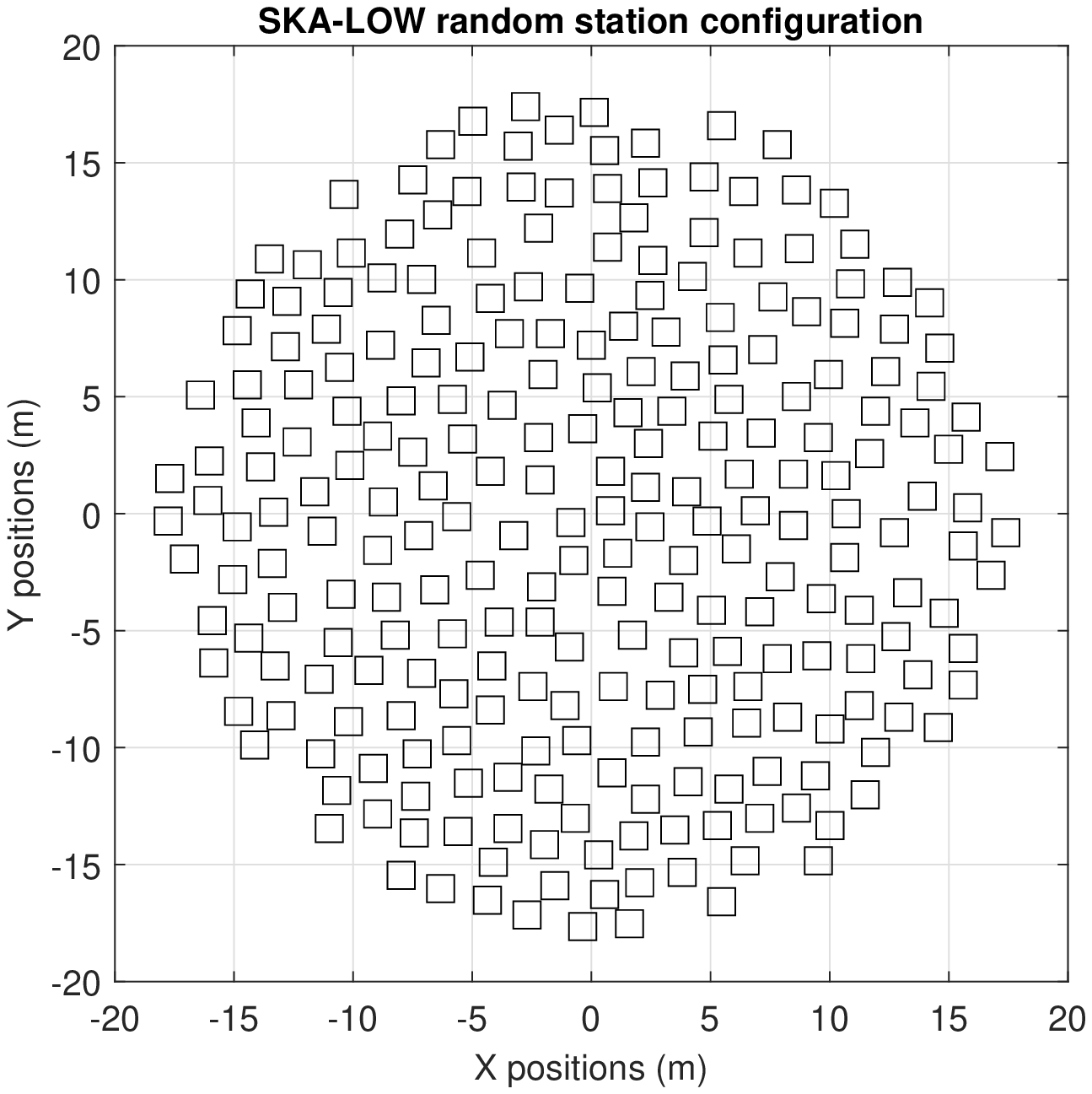}
\includegraphics[width=3.3in]{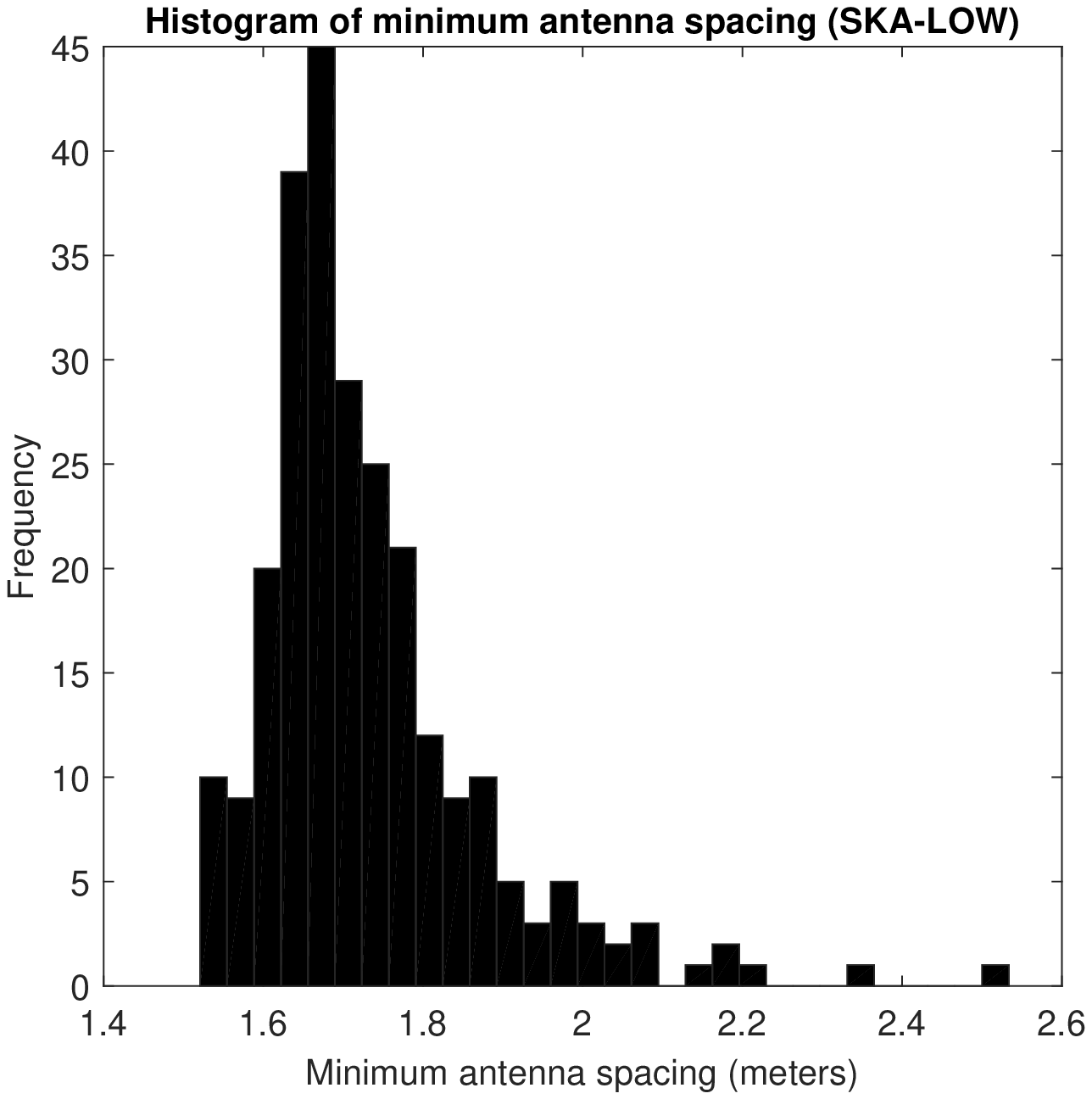}
\caption{An example of an SKA-LOW random station (left) and the corresponding histogram of minimum antenna spacing (right).}
\label{fig2}
\end{figure*}

The placement of antennas in a pseudo-random array can be achieved by defining a minimum distance parameter, $d_{min}$, and placing antennas randomly in a defined perimeter, excluding any which violate this distance to any surrounding antennas.  It is also clear that $d_{min}$ relates to the desired filing factor for a given array size.  The smaller this number is, given the physical dimensions of the antenna, the more randomised the antenna positions become.  This is not only beneficial from the mutual coupling perspective but also because the locations of side-lobes will become more randomised.  It is important to note, however, that since $d_{min}$ is very much subject to the physical size of the antenna, the full benefits of randomisation cannot be explored without throwing away filling factor, and therefore brightness sensitivity as noted earlier \citep{Mellema2013}.  For SKA-LOW, the average spacing between antennas is $\sim1.9$~m (35~m diameter with 256 antennas) \citep{Turner2015}.  To illustrate this, Figure~\ref{fig2} shows a typical SKA-LOW station's random configuration (given a $d_{min}$ of 1.5~m) and the corresponding histogram of the minimum distance.

The fact that the antenna positions do not appear truly randomised and the majority of antennas remain in close proximity to each other results in the formation of some side-lobes which exhibit the behaviour of grating lobes in regular arrays.  Therefore, they cannot be suppressed simply by spatial or weight tapering (except in the aforementioned coherent region of the beam). Despite this limitation, even if two configurations are generated using the same $d_{min}$ parameter, the position and size of the side-lobes are not the same for each one.

An option for the SKA-LOW array configuration it to make the histogram in Figure~\ref{fig2} as broad as possible.  This can achieved by making the $d_{min}$ parameter itself a random spacing within some limit, such as 1.5~m to 2.5~m.  However, this will result in a reduced filling factor and is therefore not ideal for SKA-LOW.  This is also described in \citet{Grainge2014} by assuming that a regular array has its elements perturbed in some random fashion in order to smear out the grating lobes.

Another option is to optimise the positions of the antennas in the array in order to achieve a desired side-lobe response as studied in \citet{Cohanim2004,Kogan2000}.  In \citet{Clavier2014}, a method of optimising the positions of the antennas was presented based on three crucial steps.  The first was regarded as a spatial tapering step, which aimed to morph a desired configuration into one which could achieve the appropriate beamwidth on the sky, similar to amplitude tapering.  The second and third steps moved the antennas locally and individually in order to achieve a desired side-lobe profile, all the time maintaining the minimum distance criteria, such that no antennas were placed closer than the physical limit.  The aforementioned strategy noticeably improved as more degrees of freedom were given in the placement of antennas as previously mentioned. However, with a requirement of a close-packed array, which was Nyquist sampled at the bottom of the SKA-LOW band, this method would only achieve a modest level of improvement (naturally near the main beam).  Furthermore, the benefits of averaging out the effects of mutual coupling as would be the case in a pseudo-random array \citep[see][]{Gonzalez2011} would no longer apply.





\subsection{Apodisation schemes}
\label{sec:apod}
\begin{figure*}
\centering
\includegraphics[width=7.0in]{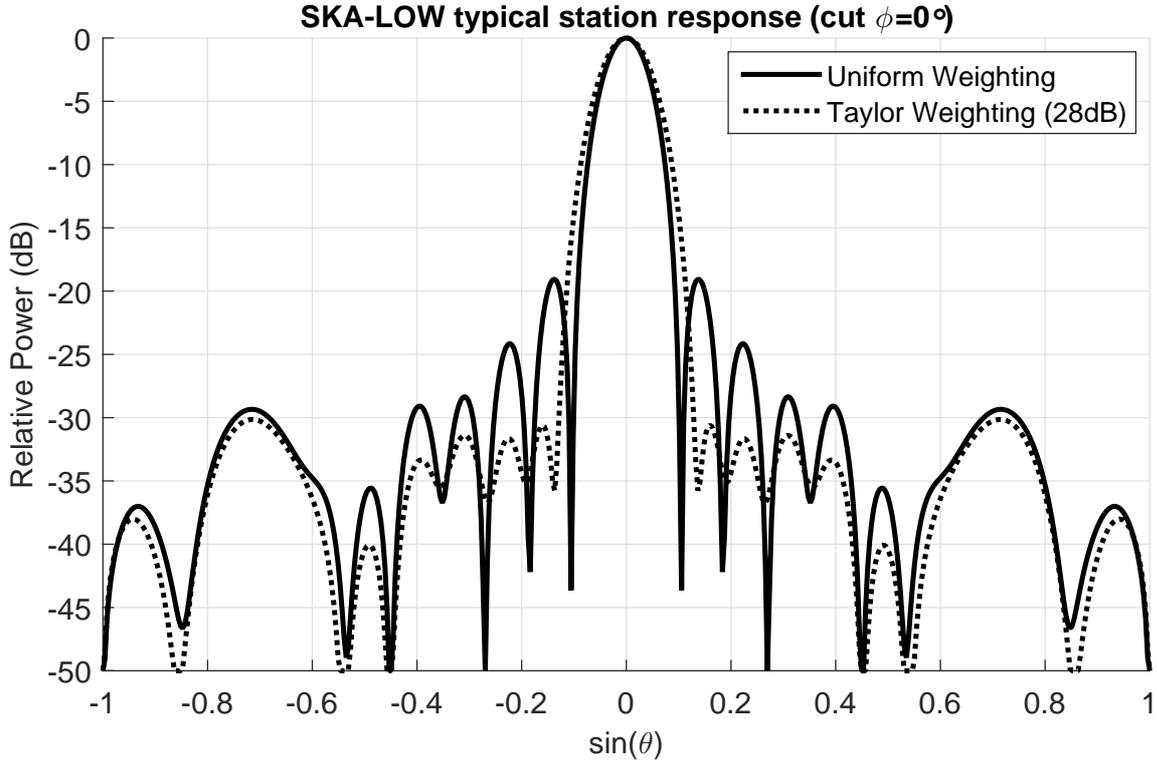}
\caption{Typical SKA-LOW station response using a Taylor (28~dB) tapering method.}
\label{fig3}
\end{figure*}

One of the major advantages of employing phased arrays is the ability to electronically control the weights applied to the signals from each antenna, thus allowing the array's main beam and side-lobe profile to be modified.  While in the strictest sense, apodisation will degrade sensitivity, as is shown in this paper for a large interferometric telescope such as SKA-LOW, it will also reduce the effects of picking up unwanted bright sources in the near-in or intermediate side-lobes. Array apodisation techniques are very common in various fields including radar and remote sensing. An example of a Taylor \citep{Taylor1955} tapering function applied to the SKA-LOW station beam (assuming randomised positions) is shown in Figure~\ref{fig3}.

For SKA-LOW, we empirically estimate the reduction in sensitivity, $S$, given by a typical tapering method to be

\begin{equation}
S(SLL_{dB}) \approx 1.47 \cdot SLL_{dB} + 26 [\%]
\end{equation}

For the example shown in Figure~\ref{fig3}, in order to reduce the first side-lobe to a level of -28~dB (azimuthally), thus providing a $>$10 dB improvement, the reduction in sensitivity is expected to be ~15\%.  Irrespective of such a modest reduction in sensitivity, such schemes can be vital to achieve the best interferometric performance for large telescopes such as SKA-LOW.  This type of tapering is one which is analysed in this paper though end-to-end simulations.

Another method, described in \citet{Buisson2015}, adjusts the amplitudes of the elements to improve the array beam when working in the dense/sparse transition region.  In \citet{Buisson2015}, the irregular array is first approximated as a continuous aperture, with one of many well-known amplitude tapering patterns applied (e.g. Taylor, Gaussian).  The elements are then treated as sampling the aperture and their amplitudes determined by the product of the amplitude of the continuous aperture at their location with an area factor. This area factor is related to the area occupied by each element. A number of definitions of occupied area may be used, including products of distances to nearest neighbours and the Voronoi cell division of the array.  Such a technique is shown to provide improvements over standard tapering methods.

In \citet{Grainge2014}, there is a proposal for SKA stations in the core to be designed such that the station beam is not only comprised of beamforming 256 antennas but rather antennas out to $\sqrt{3} r_{\mathrm{full}}$, where $r_{\mathrm{full}}$ is the baseline station radius of 17.5m for SKA-LOW \citep{Turner2015}.  This implies using three times the number of antennas (768 elements) to form the station beam and using apodisation to gain better control of the side-lobes, particularly near the main beam. Such a method of apodising overlapping stations is also discussed in \citet{Razavi2014}. Whilst this method implies losing a fraction of baselines, it does have a number of advantages, the main one being that the primary beam can have the same beamwidth as the unapodised (non-overlapping) stations and therefore will result in very little loss in sensitivity.  Furthermore, as described previously, the average far-out side-lobes are approximately given by $1/N$, where $N$ is the number of antennas in a station, meaning a further 5~dB improvement can be achieved in the average level of side-lobes. In general, there is considerable choice over the weighting function in this scheme.  One could even envisage keeping the beamwidth of the station response constant over a range of frequencies, for example in the EoR band.

Apodising the station beam by re-weighting the antennas during beam-forming has a similar effect to the visibility weighting schemes widely used in interferometric imaging, such as uniform, natural, Briggs \citep{Briggs1995} or adaptive \citep{Yatawatta2014} weighting. All of these schemes, whether applied to the station beam or the visibilities, trade sensitivity for sidelobe performance, but are commonly used to exploit data to their fullest.

%
\section{Interferometer Simulations}
\label{sec:sims}
In order to understand how the side-lobes of a phased array affect the imaging capability of the SKA-LOW telescope, a number of simulations were carried out using the OSKAR simulator to assess the FSSN metric described in Section~\ref{sec:intro}. Significant effort was put into the simulation setup so that the results would be representative of the real instrument. The main simulation parameters included the telescope model and the sky model, which are described below.

\subsection{Telescope Model}
\label{sec:telescope_model}
One of the main simulation parameters was the telescope model. Since no reference layout was available for the SKA-LOW telescope at the time of this study, a layout was generated to meet the requirements described in the SKA Level 1 Requirements Document \citep{Turner2015}.

\begin{figure*}
\centering
\includegraphics[width=3.4in]{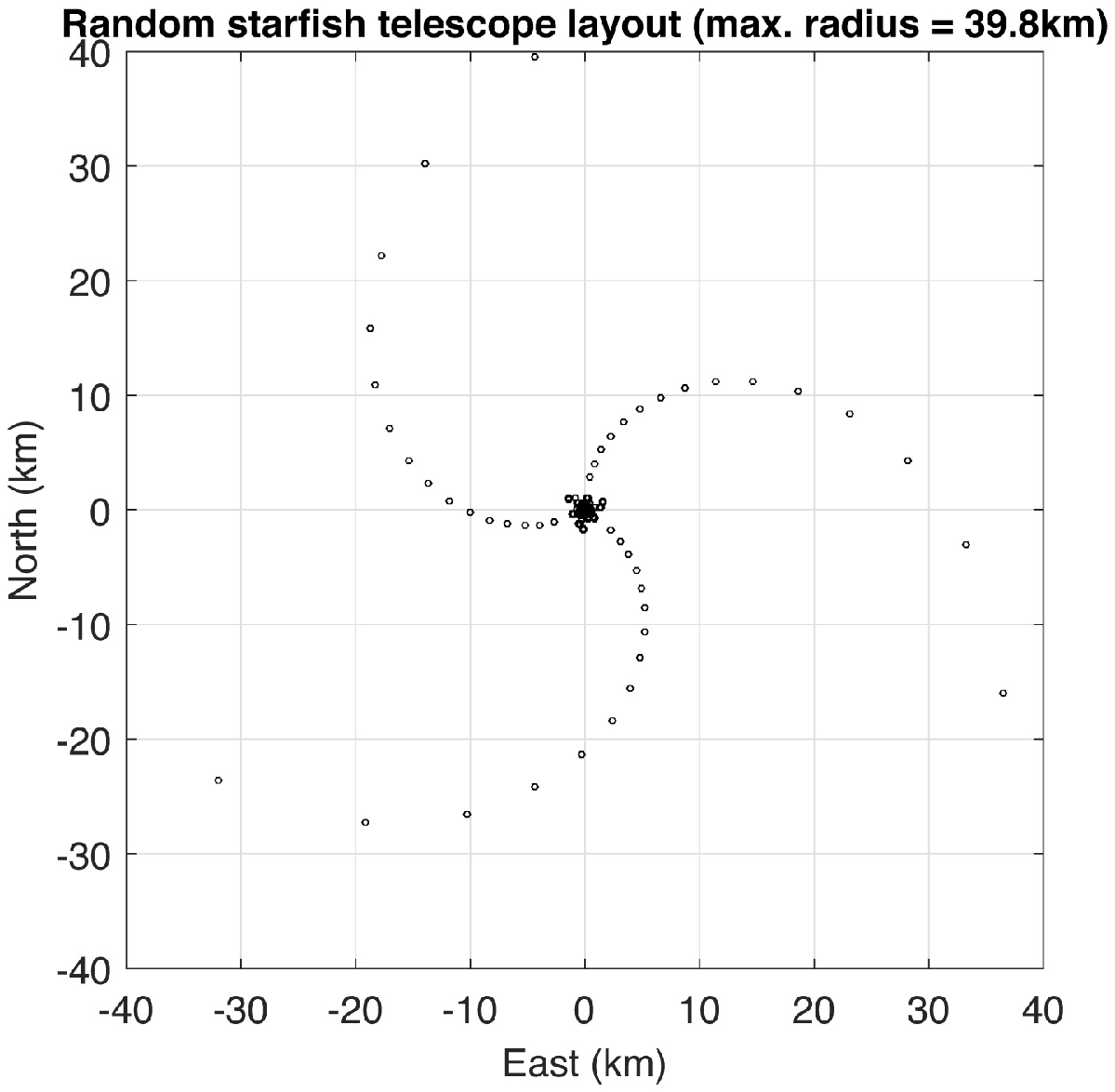}
\includegraphics[width=3.4in]{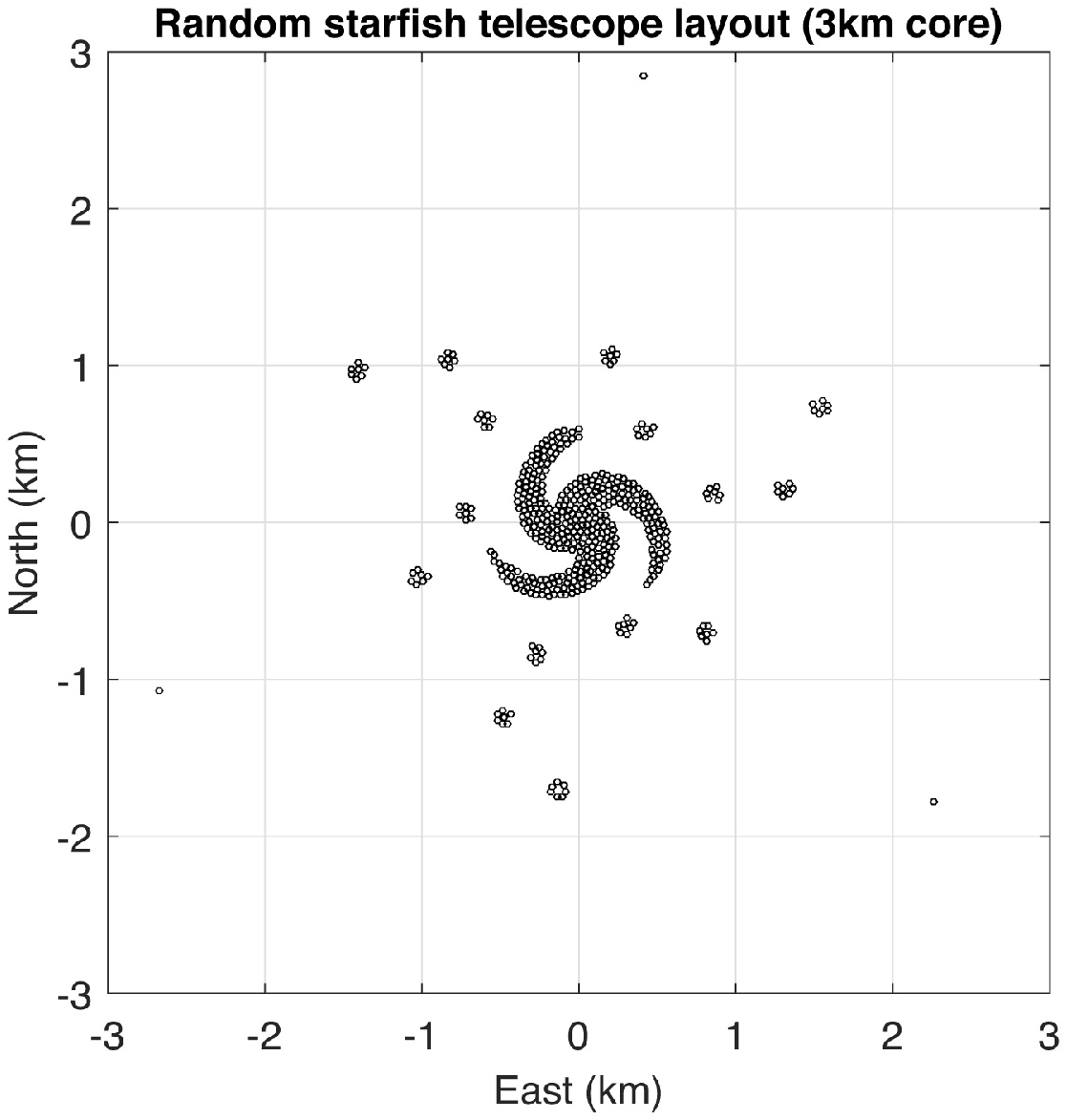}
\caption{An example telescope layout for SKA-LOW (left), and the core (right).}
\label{fig4}
\end{figure*}

The layout used for our simulations contained 512 stations in total, with 470 stations within a 3 km radius of the centre of the array as shown in Figure~\ref{fig4}.  This layout was based on a star-fish design described in \citet{Grainge2014} with 3 spiral arms.  Whilst such a layout will not provide the most optimal PSF or instantaneous UV coverage, it is a realistic option from the point of view of implementation cost.

For the intra-station configuration, a different pseudo-random layout of 256 antennas was generated for each station, with the expectation that such an approach would minimise the effects of far-out side-lobes picking up unwanted bright sources in an interferometric sense as described in Section~\ref{sec:designs}. Furthermore, our simulations included apodisation of antenna weights using a 28~dB Taylor tapering function to change the characteristics of the station beams for comparison with the un-apodised case.

For the experiments that require the highest sensitivity over large angular scales (e.g. EoR experiments), the longer baselines in the configuration shown in Figure~\ref{fig4} are likely to be used primarily for point source removal. For these simulations, a representative point source foreground would be removed as described in Section~\ref{sec:sky_model}, therefore there was justification in excluding the long baselines from this study. As such, the telescope model used in these simulations only included the 470 stations within 3 km of the core of the telescope.

In all our interferometer simulations, we used the patterns of the SKA Log-periodic Antenna (SKALA) \citep{deLera2015} generated by full-wave simulations in the CST software package to evaluate the direction-dependent antenna response at the frequencies used for this study. Detailed electromagnetic studies of array antennas in randomised configurations \citep{Gonzalez2011,deLera2011b} show that the average embedded element patterns tend towards the isolated antenna pattern, and that the better the randomisation of the antenna positions, the closer the agreement between the two. As discussed in Section~\ref{sec:designs}, this benefit is not seen with any type of periodic arrays, and even some aperiodic configurations.

\begin{figure*}
\centering
\includegraphics[width=6.0in]{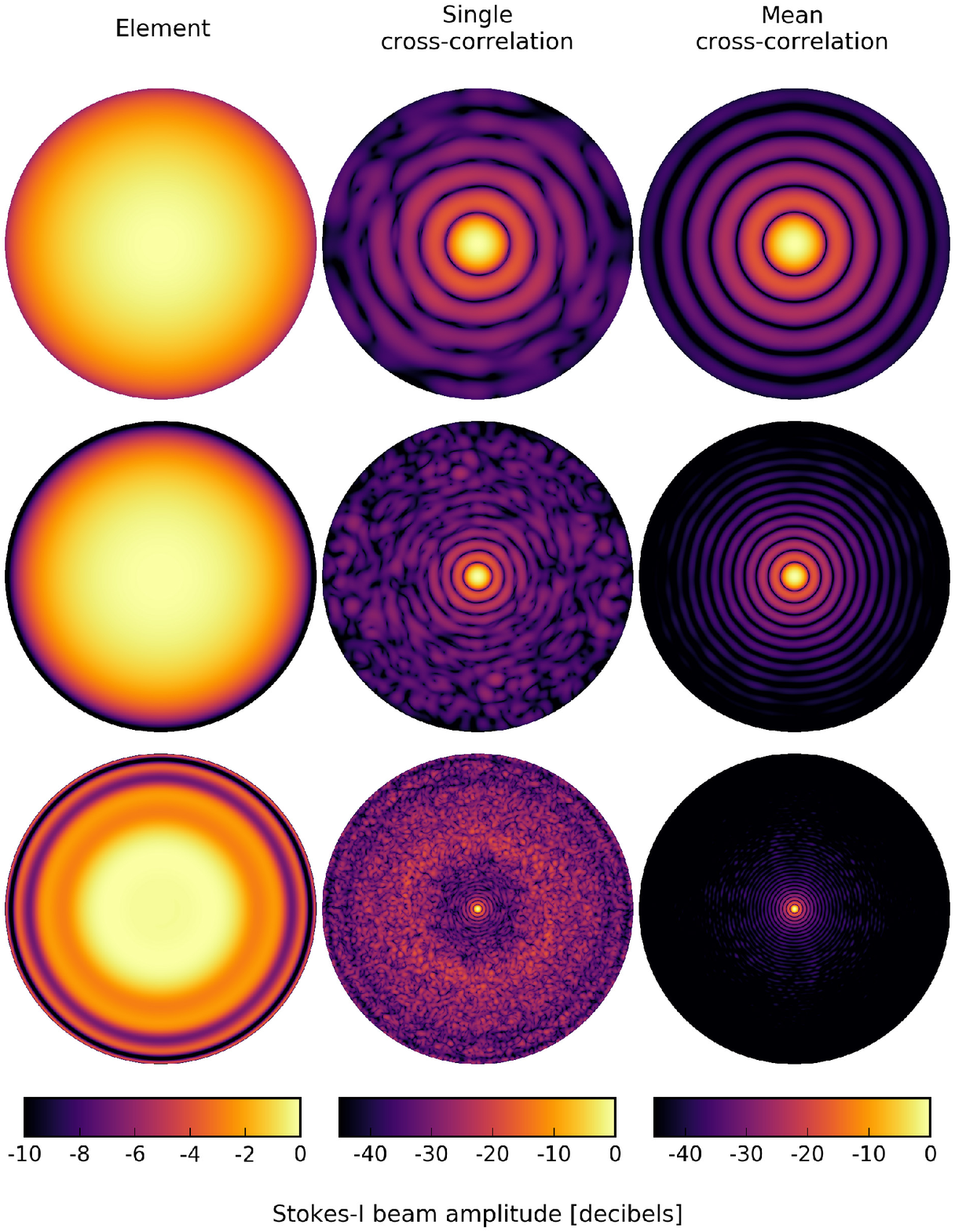}
\caption{Plots showing the all-sky coverage of the Stokes-I element beam (left column), the cross-power station beam for a single baseline (middle column), and the average cross-power station beam for the whole instrument (right column) at 50~MHz (top row), 110~MHz (middle row) and 350~MHz (bottom row).}
\label{fig5}
\end{figure*}

Plots showing the all-sky element responses in total intensity as well as images of the snapshot and time-averaged cross-power beams are shown in Figure~\ref{fig5}.

To estimate the thermal noise for this telescope, which was used for comparison with the FSSN, $A_{\rm eff}$ and $T_{\rm sys}$ were computed from full-wave array simulations, and included the clipping of effective aperture due to mutual coupling.  The description of the calculations is given in \citet{deLera2015} and \citet{Cortes1995}.  The sensitivity per element, on average, is given by

\begin{equation}
\left. \frac{A_{\rm eff}}{T_{\rm sys}} \right|_{\theta,\phi} =
\frac{\frac{\lambda^2}{4\pi} G_{\theta,\phi}}  {\eta_{\rm rad} T_{\rm A} + (1 - \eta_{\rm rad}) T_0 + T_{\rm rec}}
\end{equation}

\noindent where $A_{\rm eff}$ is the effective aperture, $T_{\rm sys}$ is the system temperature, $\lambda$ is the wavelength, $G$ is the gain of the embedded element in the array environment, $\eta_{\rm rad}$ is the radiation efficiency.  Here $T_{\rm A}$, $T_0$ and $T_{\rm rec}$ represent the antenna temperature, the ambient surrounding temperature (assumed 295K) and the receiver temperature, respectively.  At the SKA-LOW frequencies $T_{\rm A}$ is dominated by the galactic synchrotron emission and follows the empirical law $T_{\rm sky} \approx 60 \lambda^{2.55}$ \citep[e.g.][]{Turner2015}, while $T_{\rm rec}$ is dominated by the matching between the low noise amplifier and the antenna.  For the entire SKA-LOW band, the system is dominated by the sky noise.

\subsection{Sky Model}
\label{sec:sky_model}
Since FSSN is a function of the position and brightness of sources in the sky, particular care was taken to use a representative sky model that would not strongly bias the results. Because we wanted to focus our attention on discrete sources only, our sky model did not include any diffuse emission from the Galaxy.

To evaluate the response from sources in the far side-lobes of the station beam, the sky model must fully populate the field of view of the antenna elements for the duration of each simulated observation.  Since no catalogues of the full southern hemisphere at the frequency of SKA-LOW were available when these simulations were carried out, we chose to use the re-reduced VLA Low-frequency Sky Survey (VLSSr) catalogue, version 2013-08-26 \citep{Lane2012} as the basis for our all-sky model. VLSSr provides a larger total source count and improved source flux estimates over the original VLSS catalogue \citep{Cohen2007}. The survey covers the sky at declinations north of -30 degrees, at 74 MHz.  In future simulations, we intend to make use of the MWA GLEAM source catalogue of the southern sky \citep{Wayth2015}.

Any processing pipeline used for future SKA data is likely to need to remove emission carefully from the brightest sources, regardless of how far those sources are from the target field. In the LOFAR pipeline, this stage is called ``demixing'' \citep{vanderTol2007}. To simulate a perfect demixing process in the simplest way, we removed the brightest sources from the VLSSr catalogue prior to the simulations. Sources with large angular size and significant structure are represented in both VLSS and VLSSr using multiple components, all of which must be removed during demixing. By searching the original VLSS catalogue data for groups of source components that overlap on the sky, \citet{Helmbolt2008} identified 388 sources with peak fluxes $>$ 15 Jy/beam at 74 MHz, and presented radio frequency spectra for each. To find these groups, the boundary of each component was treated as an ellipse with the same position and position angle as that given by the Gaussian fit, and extending 3-sigma from the centre. Components with boundaries that intersected were considered to be overlapping, so the peak flux of the source was then determined to be the component in the group with the largest peak flux. We repeated the procedure of \citet{Helmbolt2008} but using the VLSSr catalogue data, and with 5-sigma component widths rather than 3-sigma to ensure that no components of bright sources would be omitted.

After this process, the sky model contained 92098 source components. In total, 598 components were removed from 424 bright sources with peak flux > 15 Jy/beam in the VLSSr catalogue. Sources that were listed with only an upper bound to their deconvolved sizes were treated as point sources for this work. The total integrated flux from all components removed was $\sim60$~kJy, leaving $\sim154$~kJy at 74 MHz. For the remaining components, the minimum, maximum, mean and standard deviation of the integrated fluxes were 0.3 Jy, 36 Jy, 1.7 Jy and 1.9 Jy, respectively. A spectral index value for each component was then randomly generated assuming a Gaussian probability distribution with mean and standard deviation taken from the 388 sources listed in \citet{Helmbolt2008}. The mean spectral index was -0.92, and the standard deviation was 0.22.

\begin{figure*}
\centering
\includegraphics[width=6.0in]{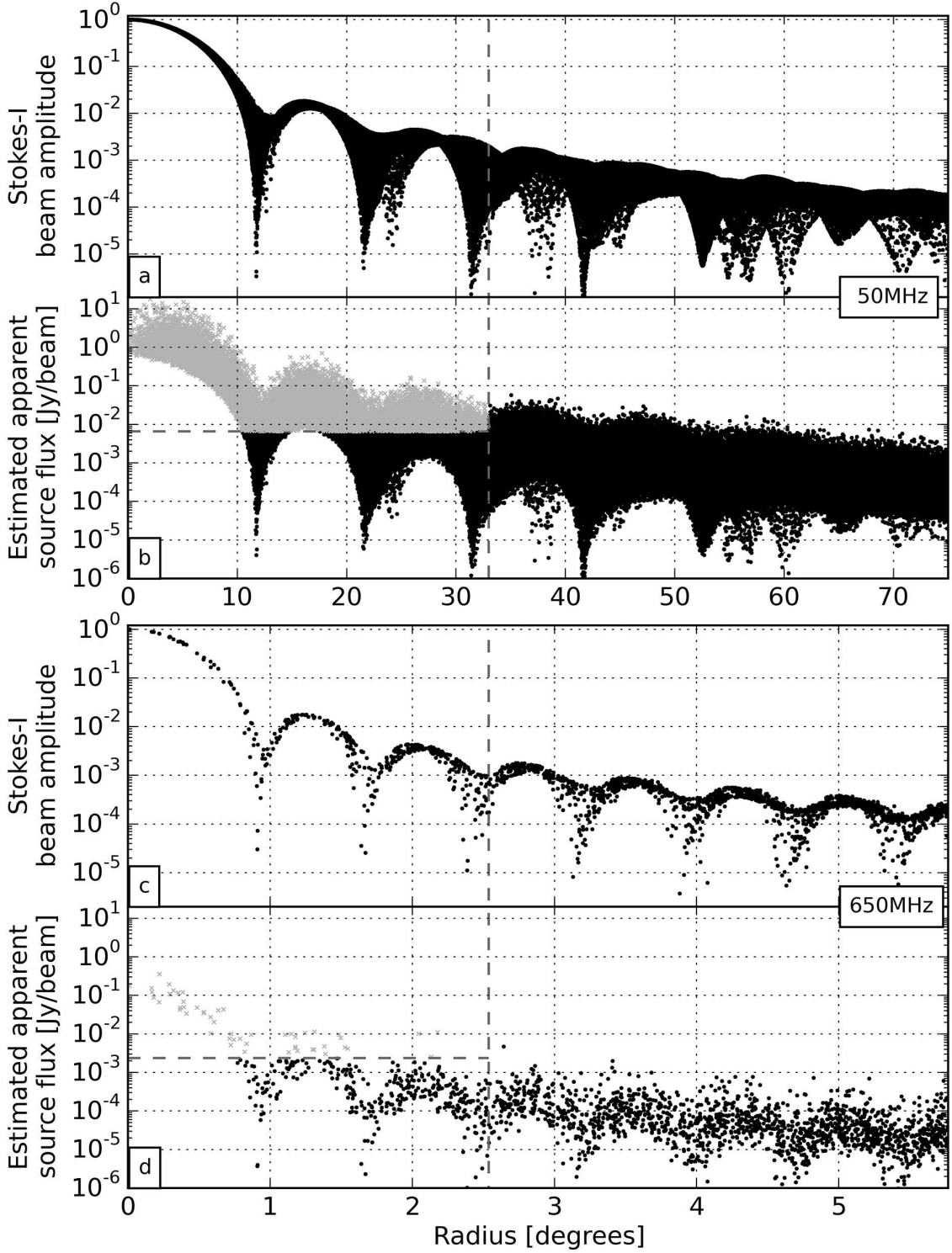}
\caption{Emulation of a perfect `CLEAN' source removal at 50 MHz (panels a \& b) and 650 MHz (panels c \& d; note the frequency-scaled x-axis values). For each simulation frequency the average 6-hour Stokes I cross-power beam of all 470 stations was evaluated at the position of all sources in the model (panels a \& c) to estimate the apparent flux of each source (panels b \& d). Sources of apparent flux above $3\sigma$ of the theoretical thermal noise for a 6-hour observation (grey points) were then removed prior to the simulation.}
\label{fig6}
\end{figure*}

In addition to the demixing process, we removed sources within the field of view out to the edge of the second side-lobe in order to emulate the CLEAN procedure that would be used to process real data. We simulated the operation of CLEAN down to a 3-sigma noise level of the telescope, based on a 6-hour observation length and appropriate bandwidth, using the levels given by models of the antenna effective area. Thermal noise values were then derived from these. Sources were removed based on their apparent fluxes, which were determined using the average cross-power Stokes I beam for the whole observation duration. Because we removed sources out to the second sidelobe, the removal radius was scaled with frequency. At 50 MHz, between 4000 and 5000 sources were removed within a radius of 33 degrees, and at 650 MHz, about 10 were removed within a radius of 2.5 degrees. Exact numbers depended on the direction of the target field. This is illustrated in Figure~\ref{fig6}.

In practice, the number of sources that can be removed will be limited by available compute power. As a result, we removed the same number of sources for the apodised station beams as that for the unapodised stations to represent the same computational cost of post-processing in both cases. Although the numbers removed were the same, precisely which sources were removed still depended on their apparent flux, and therefore also their positions in the beam.

\subsection{Observation Parameters}
\label{sec:obs_params}
In order to study the FSSN as a function of time and frequency, we selected 6 frequencies ranging over the whole extended SKA-LOW band. These were 50~MHz, 70~MHz, 110~MHz, 170~MHz, 350~MHz, and 650~MHz. The frequencies were not evenly distributed because we wanted to focus attention on the dense-sparse array transition, while also investigating the high frequency behaviour. We used observation lengths between 10 seconds and 8 hours, distributed roughly evenly in log space, with more focus at each end. Simulated observations were set up so that the phase centre would transit the meridian at the mid-point. As the sky model was relatively sparse, with a mean density of approximately 3 sources per square degree, we generated a set of 6 random target fields to ensure that the sidelobe pattern from every station would be sampled adequately. Pointing directions were chosen by ensuring that the target field did not drop below 45 degrees in elevation over 6 hours of observation time. The simulations include a thorough treatment of station beam effects by re-evaluating every station beam throughout the observation. Time and bandwidth smearing were chosen for critical (or super-critical) sampling on baselines in the core at the frequencies used for SKA-LOW, so we used a channel bandwidth of 73.2~kHz (4096 channels over the 300~MHz band) and 10.6 seconds integration time.

%
\section{Simulation results}
\label{sec:results}
To post-process the simulated visibility data, we generated dirty images using both uniform (inverse-density) and natural (unmodified) visibility weighting schemes in the WSClean software package \citep{Offringa2014}, where the image diagonal spanned the full-width-half-power of the station beam. The image sizes therefore scaled inversely with frequency. Because all the fields contained no sources out to the second sidelobe of the station beam, and because no thermal noise was added to the simulated visibilities, the RMS noise in these dirty images represents the noise floor due to sources across the sky that cannot be removed easily using traditional methods.

\subsection{Time scaling of FSSN}
\label{sec:time_scaling}
\begin{figure*}
\centering
\includegraphics[width=6.0in]{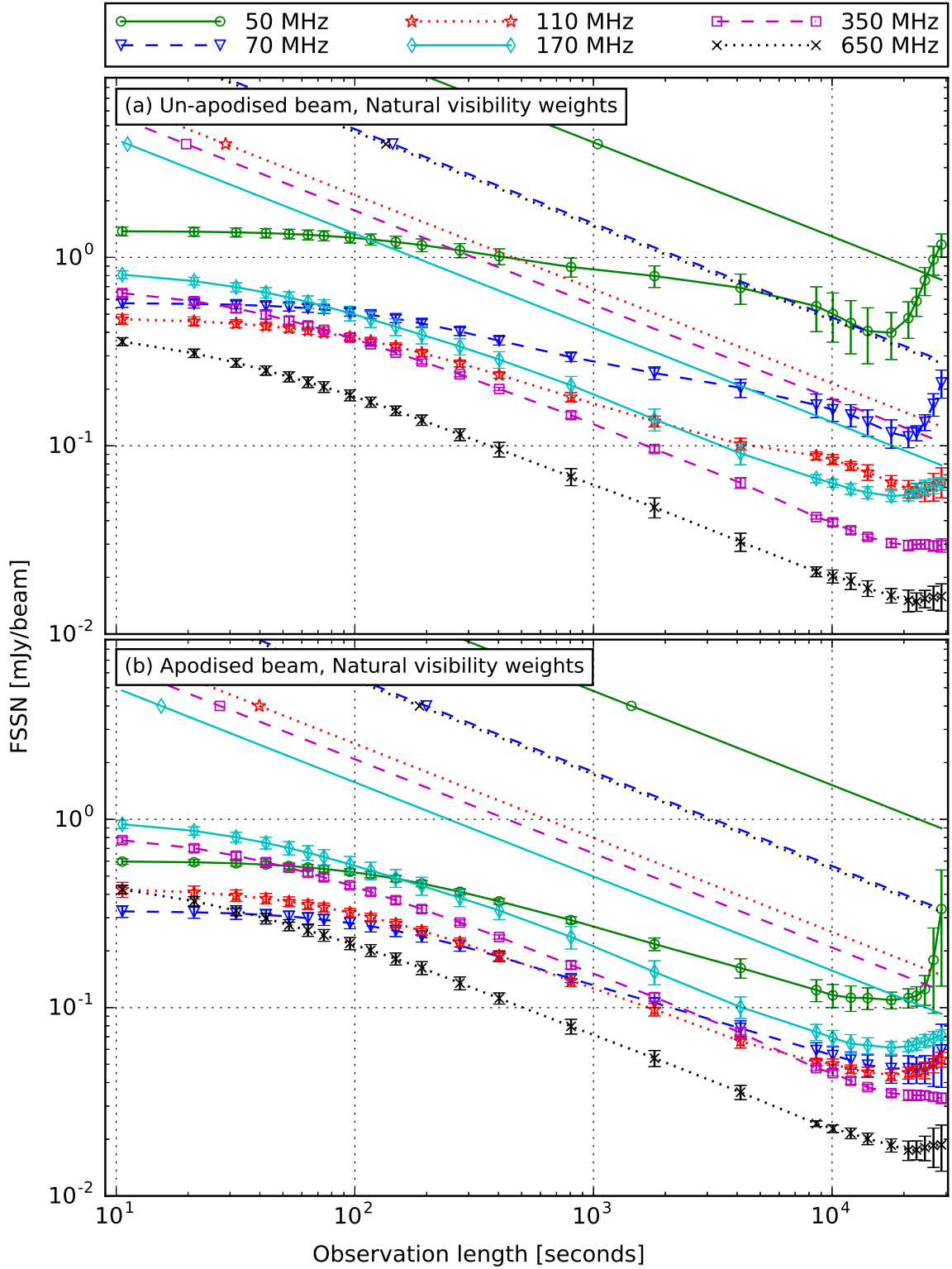}
\caption{FSSN time scaling from 10 seconds to 8 hours using natural visibility weighting for 6 observation frequencies using (a) un-apodised and (b) apodised stations. Error bars show the 1-sigma variation between 6 pointings. Straight lines represent estimated thermal noise.}
\label{fig7}
\end{figure*}

\begin{figure*}
\centering
\includegraphics[width=6.0in]{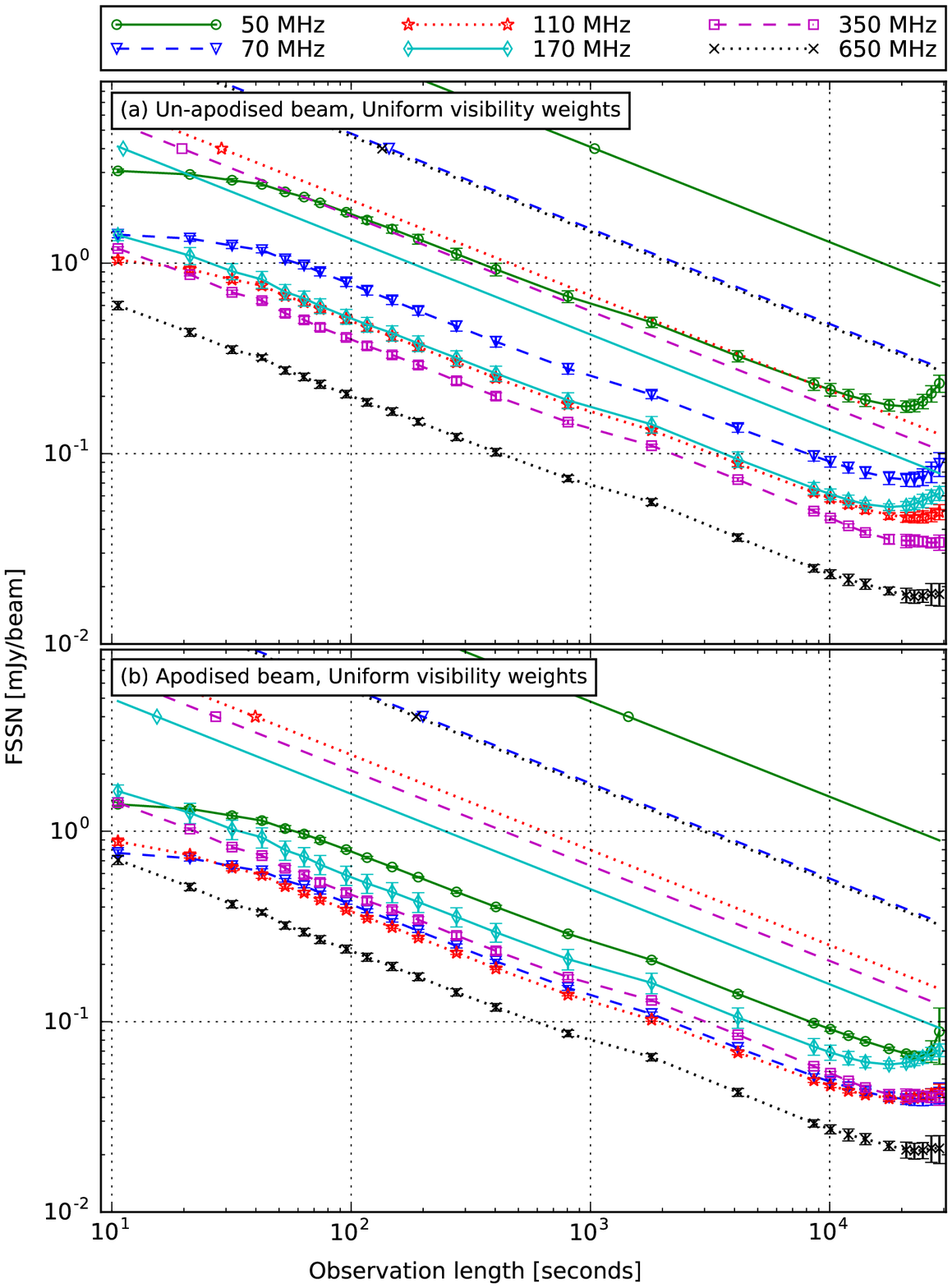}
\caption{FSSN time scaling from 10 seconds to 8 hours using uniform visibility weighting for 6 observation frequencies using (a) un-apodised and (b) apodised stations. Error bars show the 1-sigma variation between 6 pointings. Straight lines represent estimated thermal noise.}
\label{fig8}
\end{figure*}

The level of far sidelobe source noise as a function of observation length is shown in Figure~\ref{fig7} (imaged with natural visibility weights) and Figure~\ref{fig8} (imaged with uniform, or inverse-density UV weighting). The results of using un-apodised stations are shown in Figure~\ref{fig7}a and Figure~\ref{fig8}a, while the results of using station beams apodised using the 28~dB Taylor window are shown in Figure~\ref{fig7}b and Figure~\ref{fig8}b. For comparison, the straight lines on the plots show the expected thermal noise in each case, which scales as the square root of the observation time. The estimated thermal noise on the Stokes-I dirty image was evaluated using

\begin{equation}
\sigma_{\rm image,thermal} = \frac{2 k_B T_{\rm sys}}{\eta A_{\rm eff} \sqrt{4 N_b \Delta\nu \tau_{\rm obs}}}
\end{equation}

\noindent where $k_B$ is the Boltzmann constant, $T_{\rm sys}$ is the system temperature, $A_{\rm eff}$ is the average station effective area, $\eta$ is the efficiency of the station, $N_b$ is the number of baselines, $\Delta\nu$ is the channel bandwidth, and $\tau_{\rm obs}$ is the observation length.

Up to simulated observation lengths of about 6 hours, our data show that the level of FSSN generally decreases with increasing observation length. This can be understood because the synthesised aperture becomes more filled, so the side-lobes of the interferometer's point spread function become lower, and will therefore cause less power to be spilled into the target field.

After about 6 hours, the target fields move to sufficiently low elevations that the effect of the antenna element pattern becomes more significant. Because the antennas are most sensitive at the zenith, the process of amplitude calibration for station beams that are much closer to the horizon means that the power from sources far away is increased relative to those near the phase centre, so the FSSN also starts to increase at lower target elevations.

By comparing Figure~\ref{fig7} and Figure~\ref{fig8}, we see that using uniform visibility weights causes the initial overall level of FSSN to become higher than when using natural visibility weights, but also causes it to scale better with observation time. Uniform weighting makes the side-lobes of the instantaneous PSF higher than natural weighting, because uniform weighting down-weights regions of the aperture plane which are more fully sampled. The re-weighting also has the effect of increasing the resolution by increasing the relative weight of the longer baselines. This explains why the FSSN is initially higher for uniform weighting, but the smaller spatial scale of the side-lobes also means that they integrate towards zero faster: The smaller spatial scale means that summing the larger number of PSF side-lobes within the target field gives a better approximation to a normal distribution. While apodising the station beam reduces the effect of FSSN, particularly at low frequencies, the scaling behaviour with time does not change with apodisation, as it is dominated by coverage of the aperture plane.

The thermal noise level will continue to decrease with time because it is simply Gaussian random noise. Since each new data point is independent, thermal noise should always scale inversely with the square root of the integration time. However, unlike thermal noise, the level of FSSN will not change significantly once the aperture plane becomes as filled as it can possibly be. Therefore, for a fixed set of baselines, the far sidelobe source noise may well become the limiting factor to the telescope's sensitivity for an observation (or a series of observations) longer than 12~hours, which is when the PSF will no longer improve due to Earth-rotation aperture synthesis.

\begin{figure*}
\centering
\includegraphics[width=5in]{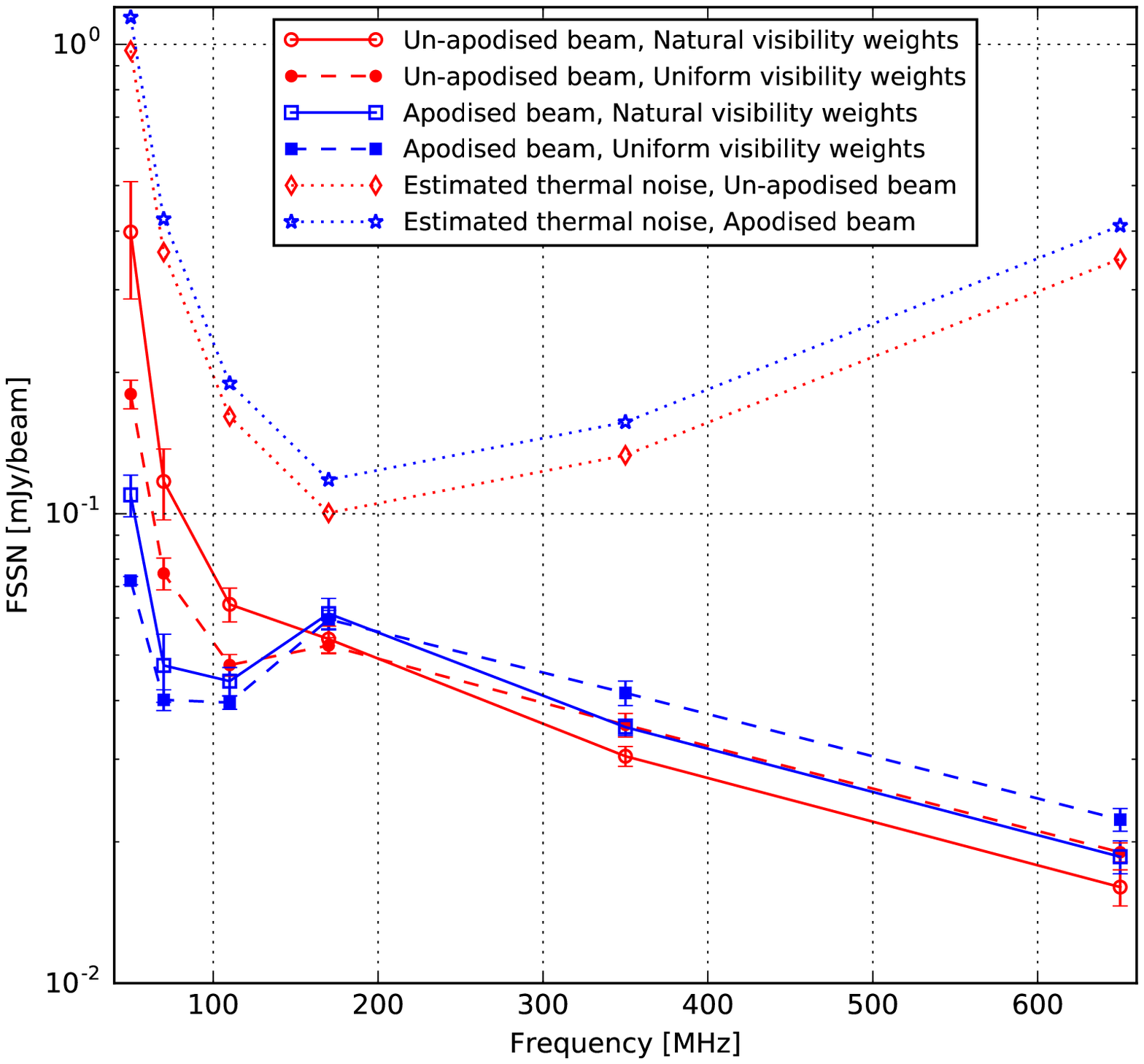}
\caption{FSSN frequency scaling for a 17702~second ($\sim 5$~hour) observation, for both unapodised station beams (circular symbols) and apodised station beams (square symbols), imaged using natural visibility weighting (solid lines/empty symbols) and uniform visibility weighting (dashed lines/filled symbols). The dotted lines show the estimated thermal noise for the observation. Error bars are drawn to show the 1-sigma variation between the 6 simulated pointings.}
\label{fig9}
\end{figure*}

At 50~MHz, the thermal noise will reach the same level as our far sidelobe source noise after only $\sim27$~hours with this telescope configuration, using an un-apodised station beam with natural visibility weighting. Whilst this may appear concerning for deep imaging experiments that would require thousands of hours of integration time to reach the required noise level (if thermal noise was assumed to be the only limit for those experiments), in practice, longer observations may also clean deeper. Although it is hard to extrapolate to a length of observation where this becomes a serious issue, our results suggest that for observations of order 100~hours, FSSN will eventually present a noise limit above the expected thermal noise. This is especially true because our simulations removed all sources in our sky model within the main lobe of the station beam, and there is a practical limit to the number of sources that can be removed for a reasonable post-processing cost.

\subsection{Frequency scaling of FSSN}
\label{sec:freq_scaling}
The level of far sidelobe source noise as a function of frequency from 50~MHz to 650~MHz, for a simulated observation length of 5~hours, is shown in Figure~\ref{fig9}. Because all the stations are very sparse at the top end of the frequency band, it was reasonable to expect that the high level of grating lobes in the station beam would spill more power into the images from sources far away, but this is not what was observed. Generally, we observe that the level of FSSN is inversely proportional to frequency, with the highest level of FSSN at 50~MHz.

The reduction in FSSN image RMS at higher frequencies can be explained by:

i) The sky becoming fainter.

ii) The primary beam (dominated by the array factor) becoming a more effective spatial filter on the sky with increasing frequency, thereby decreasing the total apparent flux of the sky.

iii) The area on the sky sampled by the first few side-lobes of the primary beam decreases with increasing frequency as the beam gets narrower.  Sources in this part of the beam make up a large contribution to the FSSN, and as the density of bright sources in the sky model remains constant, the FSSN contribution from the near in side-lobes decreases.

Figure~\ref{fig9} demonstrates that a measurable component of the noise floor in any image, i.e. the FSSN, improves as a function of frequency.  There is also a reduction in the level of FSSN when using station beam apodisation at low frequencies, although the opposite effect is observed at high frequencies.  These results highlight the challenges SKA-LOW is likely to be faced with in terms of calibration and imaging at the low end of the band, where the wider station beam requires many more sources to be removed.  Considering this metric alone, it is clear that imaging performance at 350~MHz to 650~MHz should be no worse than at lower frequencies.

%
\section{Conclusions}
\label{sec:conclusions}

As our results have shown, the FSSN is a function of both the station beam and the interferometric point spread function. While the FSSN signal is not uncorrelated noise, it does decrease as observation time increases, up to the point where the UV coverage no longer improves or antenna directionality causes problems. Apodisation of the station beam to reduce the level of the near-in side-lobes had a noticeable improvement on the level of FSSN at low frequencies, and our choice of apodisation function was able to reduce the level of FSSN by a factor of $\sim 2$ for only a 15\% loss in sensitivity. For this reason, it may be worth investigating whether frequency-dependent apodisation would be worthwhile.

The FSSN is less than the thermal noise of the telescope over a 6-hour observation, but we do not expect it to decrease at the same rate as a function of time. Improving the instantaneous aperture coverage, either by introducing more stations or adopting a less core-dominated configuration, would go some way to improving the FSSN, but this has cost implications and would reduce sensitivity to extended objects over the design used here. Without introducing more stations, another way to improve coverage of the aperture plane would be to adopt a reconfigurable station layout within the core, which allows the telescope to have a more fully filled aperture as observations continue beyond 12 hours. For instance, dynamic logical regrouping of the elements into different stations could not only improve the shape and sidelobe cancellation of the station beams, but also create different sets of baselines which could help to keep the FSSN below the thermal noise. Maintaining the flexibility to do this by utilising a ``sea'' of elements in the core is therefore much more desired.

The FSSN does not indicate problems when using highly sparse arrays, which is not what one might expect. Because all stations have different randomised layouts, the resulting cross-power beam suppresses the far-out side-lobes of the station beam very well; and at higher frequencies, the main lobe of the station beam gets smaller and therefore acts as a better spatial filter on the sky. We note, however, that our sky model did not include diffuse emission from the Galaxy, and did not include sources down to a low flux limit, so the sampling of the station beam side-lobes would have been worse at high frequencies as the beam became smaller. Future work using GLEAM \citep{Wayth2015}, which has a higher source count, would help to address this.

The effects of the ionosphere will act to reduce FSSN by smearing sources as a function of time, but these effects were not included in this simulation. However, ionospheric effects will also make it harder to remove the brightest sources, unless the removal is done on very short timescales. Smearing out sources is equivalent to replacing the bright sources with a larger population of weak sources, but the scaling of FSSN in this regime requires further investigation.

\section*{Acknowledgements}

The authors wish to thank Chris Carilli, Mike Jones, Andy Faulkner and Paul Alexander for helpful comments during the preparation of this paper.

This work used the Wilkes GPU cluster at the University of Cambridge High Performance Computing Service (http://www.hpc.cam.ac.uk/), provided by Dell Inc., NVIDIA and Mellanox, and part funded by STFC with industrial sponsorship from Rolls Royce and Mitsubishi Heavy Industries.


\bibliographystyle{mnras}
\bibliography{paper_mnras.bbl}

\bsp	
\label{lastpage}
\end{document}